\documentclass{PoS}

\title{Universality of phase diagrams in QCD and QCD-like theories}

\ShortTitle{Universality of phase diagrams in QCD and QCD-like theories}

\author{Masanori Hanada$^{\dagger}$
         \\
        Department of Physics, University of Washington, Seattle, WA 98195-1560, USA\\
        E-mail: \email{mhanada@u.washington.edu}}
        
\author{\speaker{Naoki Yamamoto}%
         \thanks{Supported by JSPS.}\\
        Institute for Nuclear Theory, University of Washington, Seattle, WA 98195-1550, USA\\
        E-mail: \email{nyama@u.washington.edu}}

\abstract{We show the universality of phase diagrams in QCD and QCD-like 
theories through the large-$N_c$ equivalence.
The whole phase diagrams are identical between QCD at finite isospin chemical potential
and SO$(2N_c)$ and Sp$(2N_c)$ gauge theories at finite baryon chemical potential.
In the chiral limit, they are also identical to that of QCD at finite chiral chemical potential.
Outside the pion or diquark condensed phase in these theories, the universality 
of phase diagrams is applicable to QCD at finite baryon chemical potential. 
We further argue that the universality may work approximately even for $N_c=3$. 
Our result makes it possible to study QCD at finite baryon chemical potential 
and high temperature, especially the chiral phase transition, 
using sign-free theories on the lattice.}

\FullConference{XXIX International Symposium on Lattice Field Theory \\
July 10-16 2011\\
Squaw Valley, Lake Tahoe, California}

\begin{document}

\section{Introduction}
One of the most important questions within the standard model is to 
unravel the phases of QCD. In spite of its various phenomenological 
significance, understanding of the properties of QCD at finite baryon 
chemical potential $\mu_B$ has been hampered mainly because of the sign problem: 
the fermion determinant in the QCD action is no longer real and positive 
at $\mu_B \neq 0$ so that the conventional Monte Carlo technique fails. 
Fortunately, there are class of theories which may resemble QCD at $\mu_B \neq 0$ but 
have no sign problem even at nonzero chemical potential. 
Such examples include QCD with isospin chemical potential $\mu_I$ \cite{Son:2000xc},
two-color QCD with even degenerate flavors $N_f$ at $\mu_B \neq 0$ \cite{Kogut:1999iv,Kogut:2000ek},
QCD with fermions in the adjoint representation at $\mu_B \neq 0$ \cite{Kogut:2000ek}, 
SO$(2N_c)$ gauge theory at $\mu_B \neq 0$ \cite{Cherman:2010jj, Cherman:2011mh, Hanada:2011ju} 
(see Sec.~\ref{sec:classification}), and Sp$(2N_c)$ gauge theory
with even degenerate flavors $N_f$ at $\mu_B \neq 0$  \cite{Hanada:2011ju}.
Although one can study the properties of these sign-free theories
on the lattice, it is a priori unclear how and if 
these theories capture the physics of real QCD at $\mu_B \neq 0$.

\begin{figure}[t]
\begin{center}
\includegraphics[width=10cm]{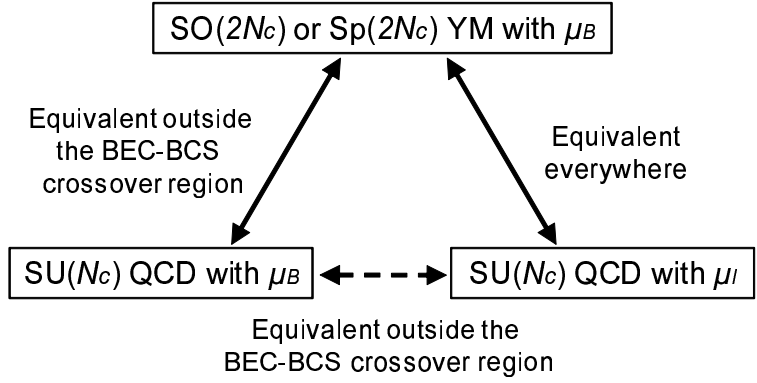}
\end{center}
\vspace{-0.5cm}
\caption{Relations between SU$(N_c)$ QCD at $\mu_B \neq 0$ and $\mu_I \neq 0$ 
and SO$(2N_c)$ and Sp$(2N_c)$ Yang-Mills (YM) theories at $\mu_B \neq 0$
through the large-$N_c$ orbifold equivalence. 
QCD at $\mu_I \neq 0$ can be obtained from SO$(2N_c)$ or 
Sp$(2N_c)$ gauge theory at $\mu_B \neq 0$ by the orbifold projection 
in the whole phase diagram, while QCD at $\mu_B \neq 0$ can be obtained 
outside the BEC-BCS crossover region. As a result, QCD at $\mu_B \neq 0$ 
is equivalent to QCD at $\mu_I \neq 0$ outside the BEC-BCS 
crossover region.}
\label{fig:QCD}
\end{figure}

Recently it has been shown by the present authors \cite{Hanada:2011ju} 
that the whole or the part of phase diagrams of these theories are 
{\it universal} if one takes the limit of large number of colors $N_c$.
The relations are summarized in Fig.~\ref{fig:QCD} (see Sec.~\ref{sec:orbifold}).
An equivalence between SO$(2N_c)$ gauge theory at $\mu_B \neq 0$ 
and QCD at $\mu_B \neq 0$ was first pointed out in \cite{Cherman:2010jj} 
and was further investigated in \cite{Cherman:2011mh}.
The universality of phase diagrams can also be extended 
(for even $N_f$ in the chiral limit)
to the other sign-free theory, QCD at nonzero chiral chemical potential 
$\mu_5$ (see Sec.~\ref{sec:chiral}),
which is of some interest in relation to the 
chiral magnetic effect \cite{Fukushima:2008xe}.
From this universality, one can learn QCD phase diagram at $\mu_B \neq 0$ 
by using {\it sign-free} QCD at $\mu_I \neq 0$ and SO$(2N_c)$ and 
Sp$(2N_c)$ gauge theories at $\mu_B \neq 0$ in the large-$N_c$ limit, 
and hopefully, for $N_c=3$. There are actually evidences 
that the universality is valid approximately even in three-color QCD
(see Sec.~\ref{sec:approximate}). 

\section{Phase diagram of QCD-like theories: an example of SO$(2N_c)$ gauge theory}
\label{sec:classification}
The Lagrangian of the gauge theories in the Euclidean spacetime is given by
\begin{eqnarray} 
\label{eq:Lagrangian}
\mathcal{L}_{G} 
=\frac{1}{4 g_{G}^{2} } {\rm Tr} (F^G_{\mu \nu})^2
+ 
\sum_{f=1}^{N_{f}}
%\left(
\bar{\psi}^G_{f} ({\cal D} + m) \psi^G_{f},
\end{eqnarray}
where ${G}$ denotes the gauge group SU$(N_c)$, SO$(2N_c)$, or Sp$(2N_c)$
and $f$ denotes the flavor index. $F^G_{\mu \nu}$ is the field strength 
of each gauge field $A^{G}_{\mu} = A^G_{\mu a} T^{G}_{a}$. 
%with $T^{G}_{a}$ being the generators of each gauge group normalized such that 
%$\tr(T^{G}_{a} T^{G}_{b}) = (1/2) \delta_{ab}$.
The Dirac fermion $\psi^G_{f}$ belongs to the fundamental representation of 
the gauge group $G$ and we take the degenerate quark mass $m_f=m$ 
for simplicity.
The Dirac operator ${\cal D}$ is defined as
${\cal D}=\gamma^{\mu} D_{\mu} + \mu \gamma^{4}$
with quark chemical potential $\mu$, and 
${\cal D}=\gamma^{\mu} D_{\mu} + \frac{1}{2}\mu_I \gamma^{4} \tau^3$
with isospin chemical potential $\mu_I = 2 \mu$ for even $N_f$.

%%%%%%%%%%%%%%%%%%%%%% 
%%%%%%%%%%%%%%%%%%%%%%%%%

As an example, let us consider the properties of SO$(2N_c)$ gauge theory at 
$\mu_B \equiv 2N_c \mu \neq 0$. 
From $(A^{\rm SO}_{\mu})^* = -A^{\rm SO}_{\mu}$, one has the relation: 
\begin{equation}
\label{eq:anti-unitary4}
C \gamma_5 {\cal D}(\mu) C\gamma_5={\cal D}(\mu)^*,
\end{equation}
where $C$ is the charge conjugation matrix.
From (\ref{eq:anti-unitary4}) and chiral symmetry $\{\gamma_5, {\cal D}\}=0$,
if $i\lambda_n$ is one of the eigenvalues of ${\cal D}$, eigenvalues appear in  
quartet $(i\lambda_n, -i\lambda_n, i\lambda_n^*, -i\lambda_n^*)$.\footnote{Note that, 
when $\lambda_n$ is real or pure imaginary, this quartet reduces to two sets of
doublets $(i\lambda_n, -i\lambda_n)$ with their eigenvectors being linearly independent
from the anti-unitary symmetry (\ref{eq:anti-unitary4}).}
Therefore, $\det[{\cal D}(\mu)+m] \geq 0$ and the Monte Carlo 
technique is available at $\mu_B \neq 0$ \cite{Cherman:2010jj, Cherman:2011mh, Hanada:2011ju}. 

When $m = \mu_B = 0$, the Lagrangian (\ref{eq:Lagrangian}) has 
the enhanced chiral symmetry U$(2N_{f})$ at the classical level
[compared with ${\rm SU}(N_{f})_{L}\times {\rm SU}(N_{f})_{R} 
\times {\rm U}(1)_{B} \times {\rm U}(1)_{A}$ in the usual three-color QCD]
owing to the anti-unitary symmetry (\ref{eq:anti-unitary4}).
At the quantum level, ${\rm U}(1)_A \subset {\rm U}(2N_{f})$ is explicitly 
broken by the axial anomaly and SU$(2N_f)$ symmetry remains.
One can indeed rewrite the fermionic part of the Lagrangian (\ref{eq:Lagrangian})
manifestly invariant under SU$(2N_f)$,
using the new variable $\Psi=(\psi_L, \sigma_2 \psi_R^*)^T$:
\begin{equation}
{\cal L}_{\rm f}=i \Psi^{\dag} \sigma_{\mu} D_{\mu} \Psi,
\end{equation}
where $\sigma_{\mu}=(-i, \sigma_k)$ with the Pauli matrices $\sigma_k$.
The chiral symmetry SU$(2N_f)$ is spontaneously broken down to 
SO$(2N_{f})$ by the formation of the chiral condensate 
$\langle \bar{\psi}{\psi} \rangle$, leading to the $2N_f^2 + N_f -1$ Nambu-Goldstone (NG) bosons 
living on the coset space ${\rm SU}(2N_{f})/{\rm SO}(2N_{f})$.
In contrast to real QCD, there are not only ${\rm U}(1)_{B}$ neutral NG modes 
with the quantum numbers $\Pi_a=\bar{\psi} \gamma_{5} P_a \psi$ 
(just like the usual pions), but also ${\rm U}(1)_{B}$ charged NG modes with the quantum numbers 
$\Sigma_S = \psi^{T} C \gamma_5 Q_S \psi$ and 
$\Sigma_S^{\dag}= \psi^{\dag} C \gamma_5 Q_S \psi^*$.
Here $P_a$ are traceless and Hermitian $N_f \times N_f$ matrices, 
$P_a=P_a^{\dag}$ ($a=1,2,\cdots,N_f^2-1$), 
and $Q_S$ are symmetric $N_f \times N_f$ matrices, $Q_S^T=Q_S$
($S=1,2,\cdots,N_f(N_f + 1)/2$), in the flavor space.
The chiral perturbation theory describing these NG modes 
for small $\mu$ is exactly the same as that of ${\rm SU}(N_c)$ 
gauge theory with adjoint fermions $\mu_B \neq 0$ considered in \cite{Kogut:2000ek},
because their symmetry breaking patterns 
are the same \cite{Cherman:2011mh, Hanada:2011ju}.

%%%%%%%%%%%%%%%%%%%%%% 
\begin{figure}[t]
\begin{center}
\includegraphics[width=15cm]{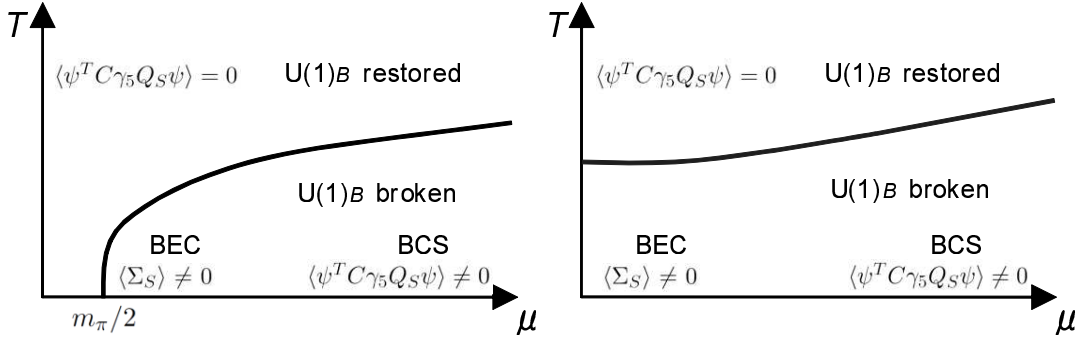}
\end{center}
\vspace{-0.5cm}
\caption{Phase diagram of SO$(2N_c)$ gauge theory at 
$\mu_B \neq 0$ for $m>0$ (left) and $m=0$ (right).}
\label{fig:SO}
\end{figure}
%%%%%%%%%%%%%%%%%%%%%%%%%

Let us consider the zero-temperature ($T=0$) ground state of the theory. 
For small $\mu> m_{\pi}/2$,\footnote{Note that the chiral perturbation theory 
breaks down when $\mu \sim m_{\rho}/2$, where $m_{\rho}$ is the mass of the 
lowest non-NG mode (i.e., $\rho$ meson mass).} it is energetically favorable 
for the U$(1)_B$ charged NG modes $\Sigma_S$ with the excitation energy 
$m_{\pi}-2\mu$ to form the Bose-Einstein condensation (BEC).
On the other hand, at sufficiently large $\mu$, 
the one-gluon exchange interaction in the $\psi \psi$-channel
is attractive in the color symmetric channel.
According to the Bardeen-Cooper-Schrieffer (BCS) mechanism, 
this leads to the condensation of the diquark pairing.
Taking into account the Pauli principle, the BCS diquark pairing 
must be flavor symmetric, and takes the form 
$\langle \psi^{T} C \gamma_5 Q_S \psi \rangle \neq 0$.
Since this BCS pairing has the same quantum numbers and breaks the same symmetry 
as the BEC $\langle \Sigma_S \rangle \neq 0$ at small $\mu_B$, 
it is natural to assume no phase transition between the BEC and BCS regions,
similarly to the BEC-BCS crossover phenomena in ultracold Fermi gases.
The phase diagram of this theory is summarized in Fig.~\ref{fig:SO} \cite{Hanada:2011ju}. 

One can check that phase diagrams of QCD at $\mu_I \neq 0$ \cite{Son:2000xc} 
and Sp$(2N_c)$ gauge theory at $\mu_B \neq 0$ \cite{Hanada:2011ju} are qualitatively similar 
to that of SO$(2N_c)$ gauge theory at $\mu_B \neq 0$ in Fig.~\ref{fig:SO}
independently of $N_c \geq 2$ though the quantum numbers of the 
condensates are different. More remarkably, one can show that these
phase diagrams are {\it universal} in the large-$N_c$ limit, 
which we shall argue below in more detail.

\section{Large-$N_c$ orbifold equivalence}
\label{sec:orbifold}
The proposed universality can be shown by using the technique of the large-$N_c$ 
orbifold equivalence. The idea of the orbifold equivalence first originates 
from the string theory \cite{Kachru:1998ys}, 
and later it is generalized within the quantum field
theories without any reference to the string theory.

The main idea is as follows. Suppose that we have some gauge theory. We first 
choose some discrete symmetry of the theory. Here we call the original theory 
the {\it parent}. Next we eliminate all the degrees of freedom which 
are not invariant under the discrete symmetry. 
We call this procedure the {\it projection}. 
As a result of the projection we obtain a new theory, which we call the 
{\it daughter}. 
Then one can show that a class of correlation functions and observables 
are equivalent between the parent and daughter theories in the 
large-$N_c$ limit. This is the {\it large-$N_c$ orbifold equivalence}. 
The field theoretical proof to the all orders in the perturbation theory 
was given in \cite{Bershadsky:1998cb}, and nonperturbative proof in 
certain gauge theories was given in \cite{Kovtun:2003hr}.
There is a caution to use this orbifold equivalence: this equivalence breaks 
down if the discrete symmetry for the projection is broken spontaneously 
in the parent theory \cite{Kovtun:2004bz}.

As an example, let us consider the projection from SO$(2Nc)$ gauge theory 
with fundamental fermions at $\mu_B \neq 0$ (parent theory) 
to QCD at $\mu_B \neq 0$ (daughter theory). 
For an earlier work of the orbifold projection from SO$(2N_{c})$ to 
SU$(N_{c})$ gauge theories, see \cite{Unsal:2006pj}.
We choose the projection conditions for the gauge field 
$A^{\rm SO}_{\mu}$ and the fermion $\psi^{\rm SO}_a$ as
\cite{Cherman:2010jj, Cherman:2011mh, Hanada:2011ju},
\begin{eqnarray}
\label{eq:projection_baryon}
A^{\rm SO}_\mu = J_c A^{\rm SO}_\mu J_c^{-1}, \qquad
\psi^{\rm SO}_{a} = \omega (J_c)_{aa'} \psi^{\rm SO}_{a'},
\end{eqnarray}
where $J_c = -i\sigma_{2} \otimes \textbf{1}_{N_{c}}$ 
and $\omega = e^{i \pi/2}$ generate ${\mathbb Z}_4$ subgroups of 
SO$(2N_c)$ and U$(1)_B$.\footnote{Here $J_c$ is chosen such that
it satisfies the {\it regularity condition} ${\rm Tr} (J_c^n) =0$ when $J_c^n$ 
does not belong to the center of SO$(2N_c)$, i.e., $J_c^n\neq\pm\textbf{1}_{2N_c}$.
This condition is necessary for the proof of the perturbative orbifold 
equivalence \cite{Bershadsky:1998cb}.}
From these projection conditions, we obtain new gauge field and new fermion field. 
By a straightforward calculation, one can show that the resulting theory is the 
U$(N_c)$ gauge theory (which can be thought as SU$(N_c)$ gauge theory 
up to $1/N_c^2$ correction at large $N_c$) with fundamental fermions at $\mu_B \neq 0$. 
Then, due to the large-$N_c$ orbifold equivalence,
a class of order parameters, e.g., the chiral condensate, 
must be equivalent between above two theories.
However, the ${\mathbb Z}_4$ discrete symmetry used for the projection 
of the fermion field, which is a part of U$(1)_B$, 
is spontaneously broken down to ${\mathbb Z}_2$ inside the BEC-BCS crossover region. 
Therefore, the orbifold equivalence between these two theories is 
valid outside the BEC-BCS crossover region of SO$(2N_c)$ gauge theory at $\mu_B \neq 0$. 

One can also construct the projection from SO$(2N_c)$ gauge theory at $\mu_B \neq 0$ 
to QCD at $\mu_I \neq 0$ for even $N_f$ 
by choosing another discrete symmetry \cite{Hanada:2011ju},
\begin{equation}
\label{eq:projection_isospin}
A^{\rm SO}_\mu = J_c A^{\rm SO}_\mu J_c^{-1}, \qquad
\psi_{af}^{\rm SO} = (J_c)_{aa'} \psi_{a'f'}^{\rm SO} (J_{i}^{-1})_{f'f},
\end{equation}
where $J_{i} = - i\sigma_2 \otimes 1_{N_f/2}$ generates ${\mathbb Z}_4$ subgroup
of ${\rm SU}(2)$ isospin symmetry and the projection condition 
for the gauge field is the same as (\ref{eq:projection_baryon}). 
In this case, the isospin symmetry is unbroken including the BEC-BCS crossover 
region (when we consider the degenerate quark mass) 
so that the orbifold equivalence holds everywhere in the phase diagram.

By repeating the same argument for Sp$(2N_c)$ gauge theory at $\mu_B \neq 0$, 
we obtain the ``family tree" of QCD and QCD-like theories as shown 
in Fig.~\ref{fig:QCD} \cite{Hanada:2011ju}. 
In particular, through the equivalence with SO$(2N_c)$ or Sp($2N_c$) gauge theory, 
we obtain the equivalence between QCD at $\mu_B \neq 0$ and QCD at $\mu_I \neq 0$ 
outside the BEC-BCS crossover region. Since QCD at $\mu_I \neq 0$ 
corresponds to the theory without the complex phase of the fermion 
determinant of QCD at $\mu_B \neq 0$, it follows that the phase-quenched 
approximation is exact in the large-$N_c$ limit there. 
%A similar statement was also suggested in \cite{Cohen:2004mw} 
%without using the orbifold equivalence.

\section{QCD at nonzero chiral chemical potential}
\label{sec:chiral}
One can also show the equivalence of phase diagrams between 
QCD at $\mu_5 \neq 0$ and QCD at $\mu_I \neq 0$ in the chiral limit 
for even $N_f$. 
To see this, we first consider the orbifold projections from 
SO$(2N_c)$ gauge theory at $\mu_5 \neq 0$ to QCD at $\mu_5 \neq 0$ 
given by (\ref{eq:projection_baryon}) and 
to QCD at nonzero isospin-chiral chemical potential $\mu_I^5$
given by (\ref{eq:projection_isospin}). 
Here $\mu_5=2\mu$ corresponds to the quark chemical potential $+\mu$ 
for $\psi_R$ and $- \mu$ for $\psi_L$,
and $\mu_I^5=2\mu$ corresponds to $+ \mu$ for $u_R$ and $d_L$ 
and $- \mu$ for $u_L$ and $d_R$ when $N_f=2$.
We then note that QCD at $\mu_I^5 \neq 0$ is equivalent to QCD 
at $\mu_I \neq 0$ by relabeling $d_L \leftrightarrow u_L$ in the chiral limit.
From the large-$N_c$ orbifold equivalence, 
the whole phase diagram of QCD at $\mu_5 \neq 0$ must be thus
identical to that of QCD at $\mu_I \neq 0$ in the chiral limit 
[or SO$(2N_c)$ gauge theory at $\mu_B \neq 0$ 
in the right panel of Fig.~\ref{fig:SO}],\footnote{Note that, 
in QCD at $\mu_I \neq 0$ or SO$(2N_c)$ gauge theory at $\mu_B \neq 0$
in the chiral limit, there is {\it no} chiral condensate
but pion or diquark condensate (see, e.g., \cite{Kanazawa:2011tt}).}
where the pion condensate $\langle \bar d \gamma_5 u \rangle \neq 0$ 
(or diquark condensate $\langle \psi^{T} C \gamma_5 Q_S \psi \rangle \neq 0$) 
is replaced by the ``chiral condensate" $\langle \bar \psi_L \psi_R \rangle \neq 0$;
the BEC-BCS crossover region of $\langle \bar \psi_L \psi_R \rangle \neq 0$ 
appears as a function of $\mu_5$.
Especially, at sufficiently large $\mu_5$, the critical temperature 
of the chiral phase transition is given by the well-known BCS formula 
$T_c = (e^{\gamma}/\pi) \Delta$, where $\Delta$ is the fermion gap at large $\mu_5$
and $\gamma \approx 0.577$ is the Euler-Mascheroni constant.

It seems that recent model calculations 
\cite{Chernodub:2011fr, Ruggieri:2011xc, Gatto:2011wc}
do not capture the physics from intermediate to large $\mu_5$ 
due to the cutoff of the model. 
We also note that, once $\mu_5>0$ is turned on, 
$\langle \bar \psi_L \psi_R \rangle$ and $\langle \bar \psi_R \psi_L \rangle$
are independent variables unlike QCD at $\mu_B \neq 0$ where 
$\langle \bar \psi_L \psi_R \rangle = \langle \bar \psi_R \psi_L \rangle$. 
Physically, as $\mu_5$ increases,
$\langle \bar \psi_L \psi_R \rangle$ should become larger at $T=0$ 
since the phase space for the pairing near the Fermi surface increases,
and so does its critical temperature; 
the genuine order parameter of chiral symmetry breaking is 
$\langle \bar \psi_L \psi_R \rangle$ rather than 
the conventional $\langle \bar \psi \psi \rangle$.

\section{Approximate universality in real QCD}
\label{sec:approximate} 
Now we discuss what to extent the universality is satisfied in three-color QCD. 
First we can explicitly check the universality at sufficiently 
large $\mu_B$ and $\mu_I$ by the weak-coupling calculations. 
For example, the fermion gap in the BCS region can be computed
by solving the gap equation as
\begin{equation}
\Delta_{G} \sim \mu {\rm exp} \left(-\frac{\pi^2}{g} \alpha_G \right),
\end{equation}
where $G$ denotes the gauge group SU$(N_c)$, SO$(2N_c)$, or Sp$(2N\c)$,
$\alpha_G$ is some factor involving $N_c$ (see \cite{Hanada:2011ju} for
the expression), and we set $g_{\rm SU}=g_{\rm SO}=g_{\rm Sp} \equiv g$. 
We find the ratios of $\alpha_G$ between QCD at large $\mu_I$ and
SO$(2N_c)$ and Sp$(2N\c)$ gauge theories at large $\mu_B$ \cite{Hanada:2011ju},
\begin{equation}
\label{eq:SO/SU}
\frac{\alpha_{\rm SO}}{\alpha_{\rm SU}} 
= \sqrt{\frac{2(N_c^2-1)}{N_c(2N_c - 1)}}
= \left\{ {\begin{array}{*{20}c}
   1.033 \quad &(N_c=3) \\
   1 \quad &(N_c=\infty) \\
\end{array}} \right. , 
\end{equation}
\begin{equation}
\label{eq:Sp/SU}
\frac{\alpha_{\rm Sp}}{\alpha_{\rm SU}} 
= \sqrt{\frac{2(N_c^2-1)}{N_c(2N_c + 1)}}
= \left\{ {\begin{array}{*{20}c}
   0.873 \quad &(N_c=3) \\
   1 \quad &(N_c=\infty) \\
\end{array}} \right. .
\end{equation} 
Not only these ratios are unity in the large-$N_c$ limit as predicted by 
the orbifold equivalence, but also they are close to unity even for three-colors. 
On the other hand, the BCS gap in QCD at large $\mu_B$ vanishes,
and there is no equivalence with this theory in this region \cite{Hanada:2011ju}. 
This is not unexpected, because the discrete symmetry used for the projection
onto QCD at $\mu_B \neq 0$ is spontaneously broken inside the BEC-BCS 
crossover region of SO$(2N_c)$ and Sp$(2N_c)$ gauge theories, and 
the orbifold equivalence should break down as explained in Sec.~\ref{sec:orbifold}.

It is also possible to check the universality within effective models 
and effective theories of QCD and QCD-like theories. 
For example, one can show that phase diagrams of chiral random matrix models 
between all the universality classes are universal \cite{Hanada:2011ju}. 
An equivalence of phase diagrams between chiral unitary matrix model
at $\mu_B \neq 0$ and that at $\mu_I \neq 0$ outside the BEC-BCS crossover region 
was first pointed out in \cite{Klein:2003fy} without using the orbifold equivalence.
An equivalence between QCD and SO$(2N_c)$ gauge theory at small $\mu_B$ 
was also confirmed at the level of chiral perturbation theories \cite{Cherman:2011mh}.
The generalization to all the class of theories in Fig.~\ref{fig:QCD}
at {\it any} quark density (at $T=0$) should be possible, 
based on the effective field theories recently constructed in \cite{Kanazawa:2011tt}.

From these nontrivial tests, we expect that the universality might work 
as an approximate notion in real QCD.

\section{Conclusion}
We have discussed the universality of phase diagrams in QCD and QCD-like 
theories through the large-$N_c$ orbifold equivalence, which may be 
valid approximately in three-color QCD. 
The proposed universality provides a way to evade the sign problem
in lattice QCD simulations at $\mu_B \neq 0$, e.g., for the 
physics related to the chiral transition.
Most recently it was rigorously shown that chiral critical phenomena, 
especially the QCD critical point \cite{Stephanov:2004wx}, are ruled 
out in QCD at $\mu_B \neq 0$ where the universality holds at large $N_c$ 
\cite{Hidaka:2011jj}.
It would be still important to understand the fate of the chiral 
transition at $\mu_B \neq 0$. 
The lattice simulations in QCD at $\mu_I \neq 0$ were already performed, 
e.g., in \cite{deForcrand:2007uz, Kogut:2007mz}. 
In our opinion, further investigations in this direction should be required.

\end{document}